\documentclass[letter]{aa} 

\usepackage{graphicx}
\usepackage{txfonts}
\usepackage{wasysym}
\usepackage{siunitx}
\usepackage{txfonts}
\usepackage[colorlinks=true,linkcolor=blue,citecolor=blue]{hyperref}%
\usepackage{color}
\usepackage[hyphenbreaks]{breakurl}

\begin{document}

   \title{HD 110067 c has an aligned orbit}

   \subtitle{Measuring the Rossiter-McLaughlin effect inside a resonant multiplanetary system with ESPRESSO}

\author{J. Zak\inst{1, 2, 3}\fnmsep\and
        H. M. J. Boffin\inst{1}\and
        E. Sedaghati\inst{4}\and
        A. Bocchieri\inst{5}\and
        Q. Changeat\inst{6}\and
        A. Fukui\inst{7, 8}\and 
        A. Hatzes\inst{9}\and
        T. Hillwig\inst{10}\and
        K. Hornoch\inst{2}\and
        D. Itrich\inst{11}\and
        V. D. Ivanov\inst{1}\and
        D. Jones\inst{8, 12, 13}\and
        P. Kabath\inst{2}\and
        Y. Kawai\inst{14}\and
        L.~V. Mugnai\inst{15}\and
        F. Murgas\inst{8, 12}\and
        N. Narita\inst{7, 16}\and
        E. Palle\inst{8, 12}\and
        E. Pascale\inst{5}\and
        P. Pravec\inst{2}\and
        S. Redfield\inst{17}\and
        G. Roccetti\inst{1, 18}\and
        M. Roth\inst{3, 9}\and
        J. Srba\inst{2}\and
        Q. Tian\inst{17}\and
        A. Tsiaras\inst{19}\and
        D. Turrini\inst{20}\and
        J. P. Vignes\inst{21}
}

\institute{European Southern Observatory, Karl-Schwarzschild-str. 2, 85748 Garching, Germany\\
           \email{jiri.zak@eso.org} \and
           Astronomical Institute of the Czech Academy of Sciences, Fri\v{c}ova 298, 25165 Ond\v{r}ejov, Czech Republic \and
           Faculty of Physics and Astronomy, Friedrich-Schiller-Universität, Fürstengraben 1, 07743, Jena, Germany \and
           European Southern Observatory, Casilla 13, Vitacura, Santiago, Chile \and
           Dipartimento di Fisica, La Sapienza Università di Roma, Piazzale Aldo Moro 5, Roma, 00185, Italy \and
           European Space Agency (ESA), ESA Office, Space Telescope Science Institute (STScI), Baltimore, MD 21218, USA \and
           Komaba Institute for Science, The University of Tokyo, 3-8-1 Komaba, Meguro, Tokyo 153-8902, Japan\and
           Instituto de Astrof\'isica de Canarias, E-38205 La Laguna, Tenerife, Spain \and
           Thueringer Landessternwarte Tautenburg, Sternwarte 5, 07778 Tautenburg, Germany \and
           Department of Physics and Astronomy, Valparaiso University, Valparaiso, IN 46383, USA \and
           Steward Observatory, The University of Arizona, Tucson, AZ 85721, USA \and
           Departamento de Astrof\'isica, Universidad de La Laguna, 
           E-38206 La Laguna, Tenerife, Spain \and
           Nordic Optical Telescope, Rambla Jos\'e Ana Fern\'andez P\'erez 7, 38711, Bre\~na Baja, Spain \and                      Department of Multi-Disciplinary Sciences, Graduate School of Arts and Sciences, The University of Tokyo, 3-8-1 Komaba, Meguro, Tokyo 153-8902, Japan\and
           School of Physics and Astronomy, Cardiff University, Queens Buildings, The Parade, Cardiff, CF24 3AA, UK\and
           Astrobiology Center, 2-21-1 Osawa, Mitaka, Tokyo 181-8588, Japan\and
           Astronomy Department and Van Vleck Observatory, Wesleyan University, Middletown, CT 06459, USA \and
           Department of Physics, Ludwig-Maximilian University of Munich, Geschwister-Scholl-Platz 1, 80539, Munich, Germany \and
           Department of Physics and Astronomy, University College London, Gower Street, WC1E 6BT London, United Kingdom \and
           INAF - Osservatorio Astrofisico di Torino, via Osservatorio 20, I-10025, Pino Torinese, Italy \and
           American Association of Variable Star Observers
}

   \date{Received April XX, 2024; accepted April XX, 2024}

 
  \abstract
   {Planetary systems in mean motion resonances hold a special place among the planetary population. They allow us to study planet formation in great detail as dissipative processes are thought to have played an important role in their existence. Additionally, planetary masses in bright resonant systems may be independently measured both by radial velocities (RVs) and transit timing variations (TTVs). 
   In principle, they also allow us to quickly determine the inclination of all planets in the system, as for the system to be stable, they are likely all in coplanar orbits.
    To describe the full dynamical state of the system, we also need the stellar obliquity that provides the orbital alignment of a planet with respect to the spin of their host star and can be measured thanks to the Rossiter-McLaughlin effect.

   It was recently discovered that HD 110067 harbours a system of six sub-Neptunes in resonant chain orbits. We here analyze an ESPRESSO high-resolution spectroscopic time series of HD 110067 during the transit of planet c. 
   We find the orbit of HD 110067 c to be well aligned with sky projected obliquity $\lambda =6^{+24}_{-26}$\,deg. This result is indicative that the current architecture of the system has been reached through convergent migration without any major disruptive events. Finally, we report transit-timing variation in this system as we find a significant offset of 19 $\pm$ 4 minutes in the center of the transit compared to the published ephemeris.}


   \maketitle
%

\section{Resonant planetary systems}
Resonant planetary chains consist of three or more planets in the same system with integer period ratios. Their configuration is highly indicative of dissipative processes playing an important role during their formation \citep{liss11,kl12,pich19}. 
One of the main proposed channels of planet formation is core accretion coupled by disc migration \citep{gold80,ter07,cole16}. During their formation, planetary embryos migrate inward due to torques from the gaseous disc. Migration is halted by the planet disc-edge interaction and other embryos migrate into a resonant chain. Current simulations predict that as gas discs dissipate, only around 50-60\% of resonant chains become unstable \citep{izi17}. This is in clear contrast to the observed population, where only a few percent of \textit{Kepler} multi-planet systems are near first-order resonance \citep{fab14}. Hence, further investigations of
these few systems in resonance are highly warranted to explain why only such a small number of systems that retained their peculiar resonant architecture has so far been detected. Several mechanisms that might be responsible for breaking the resonances at an early stage have been proposed, such as overstable librations \citep{deck15}, non-tidal secular processes \citep{ham24} or changes in the migration rate \citep{kana20}. Thus the question arises how these few resonant systems managed to keep their peculiar architecture. Recently, by analyzing 11 resonant chain systems, \citet{wong24} found an empirical relation that supports the idea that the resonant chains are formed and maintained by stalling the migration of the innermost planet near the inner edge of the disc truncated by the magnetic fields of the protostar. However, further investigation is needed to assess the universality of this mechanism.

Planetary systems can end up with similar observable properties despite different formation and migration histories \citep{daw18}. By measuring the spin-orbit alignment, in addition to the eccentricity and stellar inclination, we can infer fundamental aspects of their evolutionary history \citep{cam16,mun18,rice22e}. In particular, the sky-projected orbital obliquity, which represents the angle between a planet's orbital angular momentum and its host star's spin axis, gives invaluable clues on the dynamical evolution of multi-planetary systems and their interactions with the protoplanetary disc or other bodies \citep{mat17}. 

To measure this quantity, one can use the Rossiter-McLaughlin (R-M) effect \citep{ross24,mcl}. Since the first detection of the R-M effect of a hot Jupiter \citep{quel00}, over 250 systems have had their stellar obliquity characterized \citep[e.g.,][]{alb22,zak24}. The observed population of exoplanets has a diverse distribution of orbits \citep{sie23}. Early studies had initially shown most multi-planetary systems having aligned orbits \citep{alb13}, but, since then, several multi-planetary systems have been characterized with significant misalignment \citep[e.g.][]{da19,hjo21}.
Several mechanisms have been proposed to explain misalignment in multi-planetary systems, e.g., primordial misalignment and gravitational perturbations from stellar or planetary-mass companions. \citet{este24} showed that secular perturbations of cold gas giants alone cannot account for the obliquity distribution of low-mass exoplanets.

Planetary systems with architectures in resonant equilibrium are indicative of the absence of any major dynamical rearrangements since their formation \citep{fan12}. Therefore, they provide an unparalleled opportunity to probe the primordial obliquities of planet-hosting stars, as coplanarity does not automatically produce alignment between the planetary orbital plane and the stellar spin axis. \citet{ri23} studied the distribution of the 3-D spin-orbit alignment $\psi$ of resonant and near-resonant systems and concluded that the observed distribution of the obliquity is not consistent with near-exact ($\psi < 5$ deg) alignment. Only two systems in a resonant chain with more than 4 planets have their obliquity determined so far due to their scarcity, faint host stars, and/or low amplitude of the R-M effect.
Finally, these systems offer us a comparison to our own Solar System that has an aligned plane to within seven degrees with the solar spin axis. Such comparison will help us understand the arrangement and distribution of various planetary systems as well as to understand the origin of the Solar system architecture and assess its uniqueness \citep{mis23}.

\section{HD 110067}
HD 110067 hosts six sub-Neptune planets in resonant chain orbits \citep{luq23}. The orbital periods span from 9 to 55 days. HD 110067 c is the second closest planet on
a 13-day orbit, with an equilibrium temperature of 700 K. The host star and planetary properties are reported in Table~\ref{tab:target}.

As HD 110067 is the brightest multi-planetary system with more than 4 planets, it is a prime target for future transmission spectroscopy observations with JWST. They will determine atmospheric molecular abundance ratios that are indicators of the formation and evolution of planetary systems \citep{tur21}. Such comprehensive characterisation is generally limited to gas giant exoplanets \citep{chan22} and will remain elusive for the TRAPPIST-1 system, due to the lack of thick atmospheres of the planets \citep{zie23}. Hence, HD 110067 is the best target for studying resonant chain systems using both obliquity and atmospheric characterisation in great detail.

\begin{table}[htbp]
\caption{Properties of the host star and planet c in the HD 110067 system. Values taken from \citet{luq23}.}
\vspace{-.4cm}
\label{tab:target}
\begin{center}
\begin{tabular}{@{ }l@{ }c@{ }c@{ }} 
\hline
& Parameter & Value  \\
\hline
Star & \textit{V} magnitude & 8.4  \\
& Sp. Type  & K0  \\
& $M_{\rm{s}}$ (M$_\odot$) &0.80 $\pm$ 0.04  \\
& $R_{\rm{s}}$ (R$_\odot$) &0.788 $\pm$ 0.008  \\
& $T_{\rm{eff}}$ (K) &  5266 $\pm$ 64    \\
& $v\,\rm{sin}i_*$ (km/s) & 2.5 $\pm$ 1.0     \\

Planet c & $M_{\rm{p}}$ (M$_\oplus$) & <6.3    \\
& $K_{\rm{p}}$ (m/s) & <1.55    \\
& $R_{\rm{p}}$ (R$_\oplus$)  & 2.388 $\pm$ 0.036  \\
& $R_{\rm{p}}/R_{\rm{s}}$  & 0.0278 $\pm$ 0.0003  \\
& Period (d) & 13.673694 $\pm \,2.4 \times 10^{-5}$   \\
& $\rm{T_0} - 2457000$ (d)  & 2657.4570 $\pm$  0.0007 \\
&$a$ (au)  &0.1039 $\pm$ 0.0013  \\

&$i$ (deg)  &89.687$\pm$ 0.163 \\
&$T_{\rm{eq}}$ (K) &699 $\pm$ 9  \\
\hline

\end{tabular}

\end{center}
\end{table}
\section{ESPRESSO spectroscopy observations}
\begin{figure}[ht]
\includegraphics[width=0.5\textwidth]{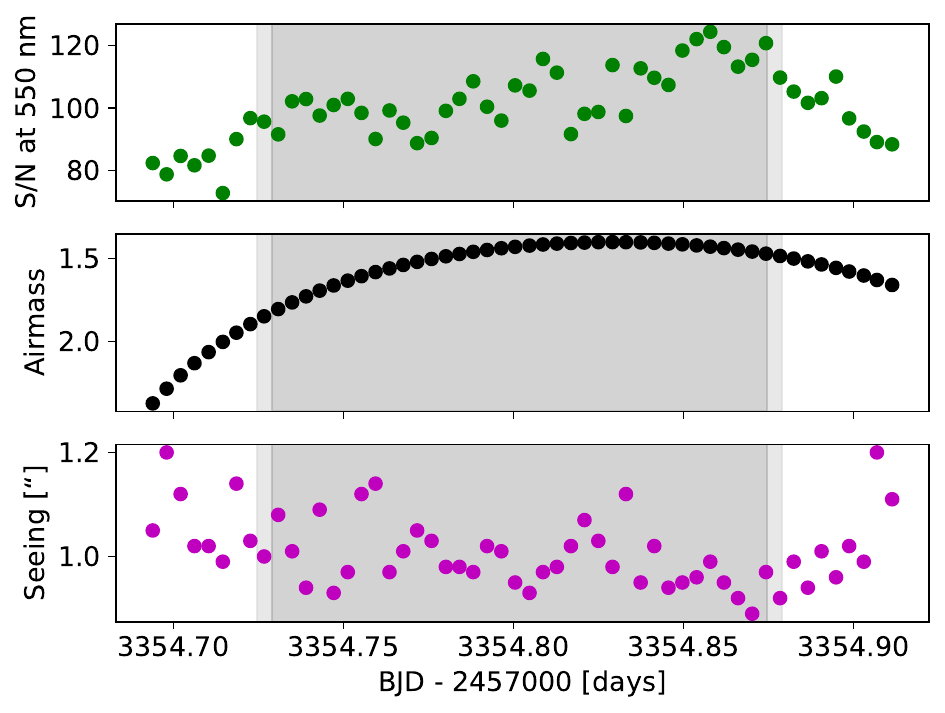}
\caption{(\textit{Top}) Signal-to-noise ratio of the ESPRESSO data at 550 nm, measured by the pipeline. (\textit{Middle}) Airmass of the target during the observations. (\textit{Bottom}) Seeing during the observations delivered at the detector and corrected for the airmass. Light grey areas indicate the partial transit duration, while dark grey shows the duration of the full transit. } 
\label{obsfig}
\end{figure}

HD 110067 was observed during the transit of planet c on 14 February 2024 with the high-resolution cross-dispersed echelle spectrograph ESPRESSO\footnote{ESO programme 112.26X6; PI: Zak} \citep{pepe21} at the Incoherent Combined Coude Focus (ICCF) of ESO's VLT at Cerro Paranal, Chile. We have used the High Resolution 1-UT mode ($\mathcal{R}\sim 140,000$), reading out the detector in the 2\,x\,1 binning setup. The readout time is 68s. The main science fiber A was placed on the target, while fiber B was used to obtain simultaneous spectra of the sky background. The instrument covers the wavelength range from 380 to 788\,nm. During the night, 53 spectra were obtained, each with an exposure time of 300s. None of the other planets in the HD 110067 system were transiting during the observations. The observation log is displayed in Table~\ref{obslog} and Fig.~\ref{obsfig}. The data were reduced using version 3.1.0 of the ESPRESSO pipeline\footnote{\url{https://ftp.eso.org/pub/dfs/pipelines/instruments/espresso/espdr-pipeline-manual-3.1.0.pdf}}. The radial velocity measurements together with their uncertainties are obtained by the ESPRESSO Data Reduction Software (DRS) pipeline, through the fitting of a Gaussian function to the cross correlation function with K0 spectral mask. Alongside the spectroscopy, simultaneous photometry was obtained, but we were not able to detect any signal from the transiting planet; the description of the photometry is presented in Appendix~\ref{simphot}.

\begin{table*}[htbp]
\caption{Observing log for the ESPRESSO data set of HD~110067. The number in parenthesis represents the number of frames taken in transit.}
\vspace{-.4cm}
\label{obslog}
\centering
\begin{center}
\begin{tabular}{lcccccc}
\hline
 Date & No.     & Exp.     & Airmass & S/N & \\ 
     &~Obs~&~Time (s)~& range  &  at 550nm  &  \\ 
\hline

2024-02-14 & 53 (37)~&~300~& 2.41-1.40-1.67 ~&~72.8-124.4 &  \\

\hline
\end{tabular}

\end{center}
\end{table*}

\section{Analysis}
\begin{figure}[ht]
\includegraphics[width=0.5\textwidth]{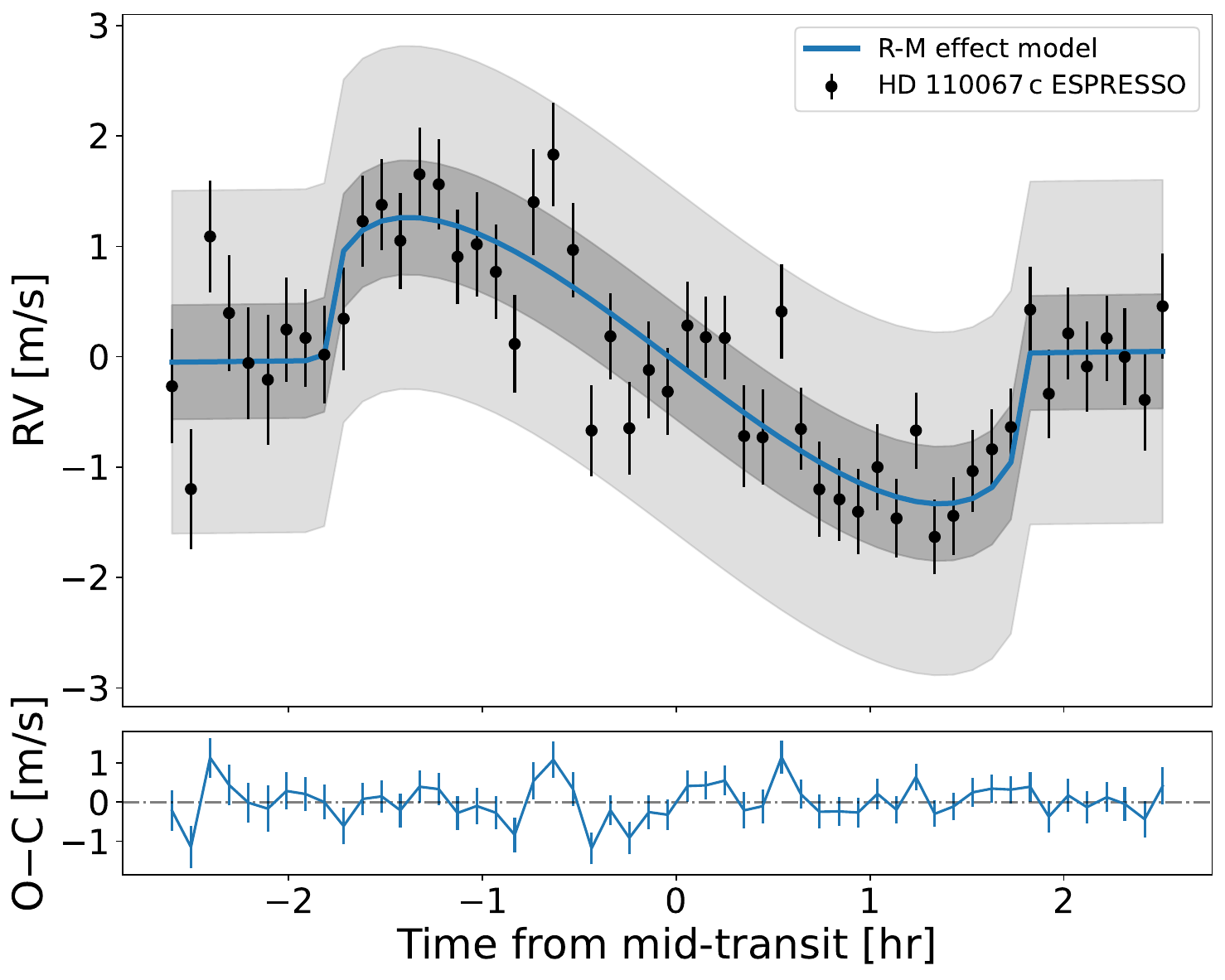}
\caption{The Rossiter-McLaughlin effect of HD 110067 c observed
with ESPRESSO. The measured data points (black) are shown
with their errorbars. The blue line shows the model that fits the data best, together with 1-$\sigma$ (dark gray) and 3-$\sigma$ (light gray) confidence intervals. The systemic velocity was removed for better visibility.}
\label{f:rm}
\end{figure}

The obtained stellar radial velocities clearly show the anomaly produced by the R-M effect (Fig.~\ref{f:rm}). To measure the projected stellar obliquity of the system ($\lambda$), we fit the radial velocities with a composite model, which includes a Keplerian orbital component, as well as the R-M anomaly. This model is implemented in the \textsc{ARoMEpy}\footnote{\url{https://github.com/esedagha/ARoMEpy}} \citep{seda23} package, which utilises the \textsc{Radvel} python module \citep{ful18} for the formulation of the Keplerian orbit. \textsc{ARoMEpy} is a Python implementation of the R-M anomaly described in the \textsc{ARoME} code \citep{bou13}. We have used the R-M effect function defined for radial velocities determined through the cross-correlation technique in our code.
We have set wide uniform priors for the radial velocity semi-amplitude $K$ and the systemic velocity $\gamma$. In the R-M effect model, we fix the following parameters to values reported in the literature: orbital period $P$ and the eccentricity $e$ to a circular orbit. The parameter $\sigma$, which is the width of the Cross-correlation function (CCF) and represents the effects of the instrumental and turbulent broadening, is measured on the data and fixed to $\sigma= 6.60$ km/s. Furthermore, we use the ExoCTK\footnote{\url{https://github.com/ExoCTK/exoctk}} tool to compute the quadratic limb-darkening coefficients with ATLAS9 model atmospheres \citep{cas03} in the wavelength range of the ESPRESSO instrument (380-788\,nm). The used limb-darkening coefficients are $u_1=0.564$ and $u_2=0.145$.  We set Gaussian priors from the discovery paper \citep{luq23} on the following parameters during the fitting procedure: central transit time $T_C$, orbital inclination $i$, the scaled semi-major axis $a/R_{\rm{s}}$ and the planet-to-star radius ratio $R_{\rm{p}}/R_{\rm{s}}$. Uniform priors were set on the projected stellar rotational velocity $\nu\,\sin i_*$ and sky-projected angle between stellar rotation axis and normal of the orbital plane $\lambda$. To obtain the best fitting values of the parameters we employed three independent Markov Chain Monte Carlo (MCMC) simulations each with 250\,000 steps, burning the first 50\,000. The results of the MCMC simulations are presented in Table~\ref{tab:res} and in Figs.~\ref{f:rm} and \ref{f:rmmcmc}. 


\begin{table}[htbp]
\caption{MCMC analysis results. $\mathcal{N}$ denotes priors with normal distribution and $\mathcal{U}$ denotes priors with uniform distribution.}
\label{tab:res}
\begin{center}
\begin{tabular}{l l  c c } 

\hline
& Parameter & Prior & Result  \\ 
\hline
 \vspace{0.05 cm}
 & T$_{\rm{c}} - 2460000 $ (d) 
 & $\mathcal{N}(T_0\pm 0.01)$ &  354.8018$^{+0.0020}_{-0.0022}$\\
 \vspace{0.05 cm}
 
 & $\lambda$ (deg) 
 & $\mathcal{U}(-180, 180)$ & 6$^{+24}_{-26}$ \\
\vspace{0.05 cm}

 & $v\,\sin{i_*}$ (km/s) 
 & $\mathcal{U}(0, 5)$ & 1.39$^{+0.26}_{-0.19}$ \\
 \vspace{0.05 cm}

 & $R_{\rm{p}}/R_{\rm{s}}$ 
 & $\mathcal{N}(0.0278, 0.0003)$ &$ 0.0278 \pm 0.0003$ \\
 
 & $a/R_{\rm{s}}$ 
 & $\mathcal{N}(28.37, 0.30)$ &  28.52$^{+0.29}_{-0.30}$ \\
 \vspace{0.05 cm}
 
 & $i$ (deg) 
 & $\mathcal{N}(89.687, 0.163)$ & 89.64$^{+0.19}_{-0.17}$ \\
 \vspace{0.05 cm}
 
 & $\gamma$ (km/s) 
 & $\mathcal{U}(-8.48, -8.58)$ &$ -8.5306 \pm 0.0001$ \\

  & $K$ (km/s) 
 & $\mathcal{U}(-0.02,0.02)$ &$ -0.001 \pm 0.003$ \\
\hline
\end{tabular}
\end{center}
\end{table}

\section{An aligned system with small TTVs}
From our analysis, we find a well aligned orbit for HD 110067~c with sky projected obliquity $\lambda =6^{+24}_{-26}$\,degrees. Such a high state of alignment is indicative of the fact that the architecture of the system has been reached through convergent migration without any major disruptive events (making the reasonable assumption that all the planets share the same value of $\lambda $). We ran a numerical simulation with one planet inclined and such system does not lead to a set of six transiting exoplanets. Given the tidal circularisation timescale of planet c \citep[$ \tau_{\rm{circ}}\gg$ 14 Gyr,][]{g66}, we can also rule out any events exciting the orbit of planet c that would later be realigned, thereby confirming the quiescent evolution of the system. 

We detected a scatter (just before the middle of the transit) in the observed RV data that we were not able to fully describe in our model, although it is not very significant (within 3-$\sigma$ interval of our model). To investigate the origin of this scatter we have checked the spectral CCF indices \citep{hat19} provided by the ESPRESSO DRS pipeline (e.g., FWHM and bisector span), as well as the $\log R^\prime_{\mathrm{HK}}$ activity index provided by the Data Analysis Software (DAS) pipeline\footnote{\url{https://www.eso.org/sci/software/pipelines/espresso-das/espresso-das-pipe-recipes.html}} that is indicative of stellar activity originating in the chromosphere. Finally, the H$\alpha$ index was measured using the Activity Indices Toolkit (ACTIN2) code\footnote{\url{https://github.com/gomesdasilva/ACTIN2}}. We do not find any correlation in any of the indices with the RV residuals. We show the derived values of $\log R^\prime_{\mathrm{HK}}$ in Fig. \ref{f:loghk} together with the RV residuals. The data show no significant deviations from the mean value throughout the night suggesting that chromospheric activity is not responsible for the scatter. The scatter can be caused by the planet crossing a stellar spot. However, despite having multiple photometry datasets we were not able to detect any features indicative of spot-crossing.

\begin{figure}[ht]
\includegraphics[width=0.5\textwidth]{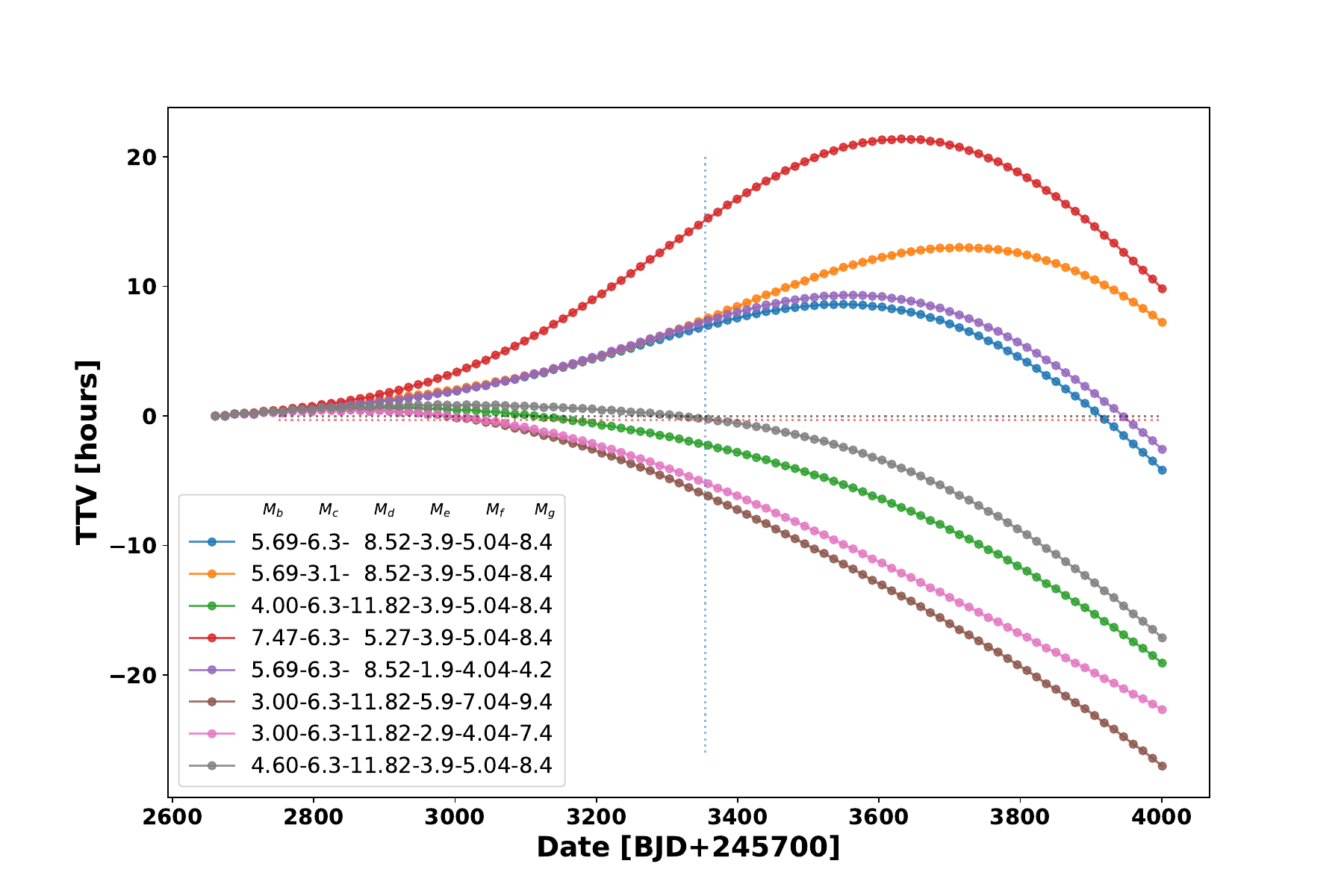}
\caption{Estimated transit time variations for planet c in the system, for various masses of the planets, as indicated in the legend. The dashed vertical line marks the observing time with ESPRESSO. The dotted horizontal lines mark no TTV (black) and a 19 min TTV (red), respectively.}
\label{f:ttv_c_mass}
\end{figure}

Our derived center of the transit is 19 $\pm$ 4 minutes earlier than what is expected from the calculated ephemeris, while the propagated uncertainty is only 3 minutes \citep{luq23}. Such significant deviation hints at transit-timing variation (TTV), which is not a surprise in a system with six planets. TTVs are observed in similar multi-planetary systems such as TRAPPIST-1, TOI-178, and TOI-1136  \citep{tey22,del23,beard24}. 

Constraining the amplitude of the TTVs is rather difficult with such a weak constraint on the planetary masses \citep{lam24}. Nevertheless, we have run numerical simulations using REBOUND \citep[]{re12} and indeed, since the published ephemeris, TTVs up to tens of hours could be expected. In this sense, having caught the transit with such a small TTV appears quite surprising, even if this is consistent with the reported nondetection of TTVs in two years of TESS observations \citep{luq23}. The masses in the system are not well known, however -- for some of them, only upper limits exist. This will lead to a rather large range of variations in the expected TTVs for planet c as shown in Fig.~\ref{f:ttv_c_mass}. Some combination of masses would explain the small amount of TTVs since the previous ephemeris. However, as the solution is clearly degenerate, more transit times are definitively needed to confirm this.
Such TTVs need to be taken into account when planning future atmospheric follow-up, during, e.g., JWST Cycle 4 or Ariel to avoid wasting valuable telescope time.

In the inside-out planet formation scenario \citep{chat14} the planets that are the closest to the host star have formed first with the ones on the farther orbit forming later. In this regard, planets farther from their host stars with weaker tidal forces offer us an even better opportunity to search for possible evidence of any perturbations. However, notable exceptions    are systems similar to K2-266 \citep{rod18}, that hosts at least 3 co-planar planets on short orbits and one ultra-short period planet ($\sim$ 0.7 days). The innermost planet with its lower inclination ($\sim 75$\,deg) significantly deviates from the orbital configuration of the other planets in the system with inclinations above 87\,deg. Such orbital configuration suggests a different evolutionary pathway for the innermost planet \citep{bec20}.

Spin-orbit measurements of multi-planetary resonant chain systems with more than 4 planets are very sparse with only two studied systems. TRAPPIST-1 was studied by \citet{hir20} who found the system well aligned from a joint analysis of three planets in the system. An aligned orbit was later confirmed by \citet{brad23} who derived an obliquity of the system $\lambda = 2^{+17}_{-19}$ deg. \citet{dai23} derived an aligned orbit, $\lambda = 5$ $\pm$ 5 deg, for the planet d in the sextuplet system TOI-1136. Together with the well aligned orbit of HD 110067 c, such a preliminary trend among resonant chains might support a proposition made by \citet{este24} that primordial misalignment may have a key role in triggering dynamical instabilities after gas disc dispersal.

We encourage further investigations of the HD 110067 system. Due to the larger R-M effect amplitudes of planets d, f, and g (compared to b, c, and e), these planets are amenable to future measurement of the obliquity with even higher precision, possibly allowing us to detect Solar System-like deviation from an aligned state \citep{beck05}. Such measurements will provide the (mis)-alignment between two or more planets in the resonant chain system with high confidence for the first time. Furthermore, such studies will also complement the photometric studies using TTVs to obtain precise masses of this system, independently of the RV method. This will aid in understanding the observed discrepancy in the literature between planetary masses derived using the RV and TTV methods \citep{mills17, adi24}.

\begin{acknowledgements}
The authors would like to thank the anonymous referee for
their insightful report. JZ, PK and JS acknowledge the support from GACR:22-30516K. DJ acknowledges support from the Agencia Estatal de Investigaci\'on del
Ministerio de Ciencia, Innovaci\'on y Universidades (MCIU/AEI) and the
European Regional Development Fund (ERDF) with reference
PID-2022-136653NA-I00 (DOI:10.13039/501100011033). DJ also acknowledges
support from the Agencia Estatal de Investigaci\'on del Ministerio de
Ciencia, Innovaci\'on y Universidades (MCIU/AEI) and the the European
Union NextGenerationEU/PRTR with reference CNS2023-143910
(DOI:10.13039/501100011033). A. Bocchieri is supported by the Italian Space Agency (ASI) with \textit{Ariel} grant n. 2021.5.HH.0. We acknowledge financial support from the Agencia Estatal de
Investigaci\'on of the Ministerio de Ciencia e Innovaci\'on
MCIN/AEI/10.13039/501100011033 and the ERDF “A way of making Europe”
through project PID2021-125627OB-C32, and from the Centre of Excellence
“Severo Ochoa” award to the Instituto de Astrofisica de Canarias. Work of KH and PP was supported by the project RVO:67985815. Based (in part) on data collected with the Danish
1.54-m telescope at the ESO La Silla Observatory. DI acknowledges support from collaborations and/or information exchange within NASA’s Nexus for Exoplanet System Science (NExSS) research coordination network sponsored by NASA’s Science Mission Directorate under Agreement No. 80NSSC21K0593 for the program ``Alien Earths”. This work is partly supported by JST SPRING, Grant Number JPMJSP2108.
\end{acknowledgements}

%
%

\bibliographystyle{aa}
\bibliography{aa}

\begin{appendix} 

\section{Simultaneous photometry}
\label{simphot}
Simultaneous photometry was secured on the night of the transit at various facilities: the SARA-CT Andor Ikon-L instrument at the SARA telescope \citep{keel17,keel21} and a 17-inch privately owned telescope at Cerro Pichasca in Deep Sky Chile observatory; the DFOSC instrument mounted at the ESO 1.54-m Danish telescope \citep{and95} and the E152\footnote{\url{https://www.eso.org/public/teles-instr/lasilla/152metre}} telescope at La Silla; an automated 24-inch telescope at Van Vleck Observatory at Wesleyan University in Connecticut, USA; and MuSCAT2 \citep{narita19} instrument mounted on the Telescopio Carlos Sánchez in the Canary Islands. In all cases, the data were reduced with standard procedures and we used DAOPHOT in its {\tt astropy} implementation to perform differential photometry of HD 110067 and the bright neighbouring star TYC 1448-459-1. 

From our obtained photometry, we can rule out any significant stellar activity at the level of a few ppt during the night of the transit. However, due to the small expected transit depth ($\sim$\,0.8\,ppt) of planet c and large airmass during the observations, we are unable to detect the planetary transit signal in our simultaneous photometry observations.

\section{Additional figures}

\begin{figure*}[ht]
\includegraphics[width=1\textwidth]{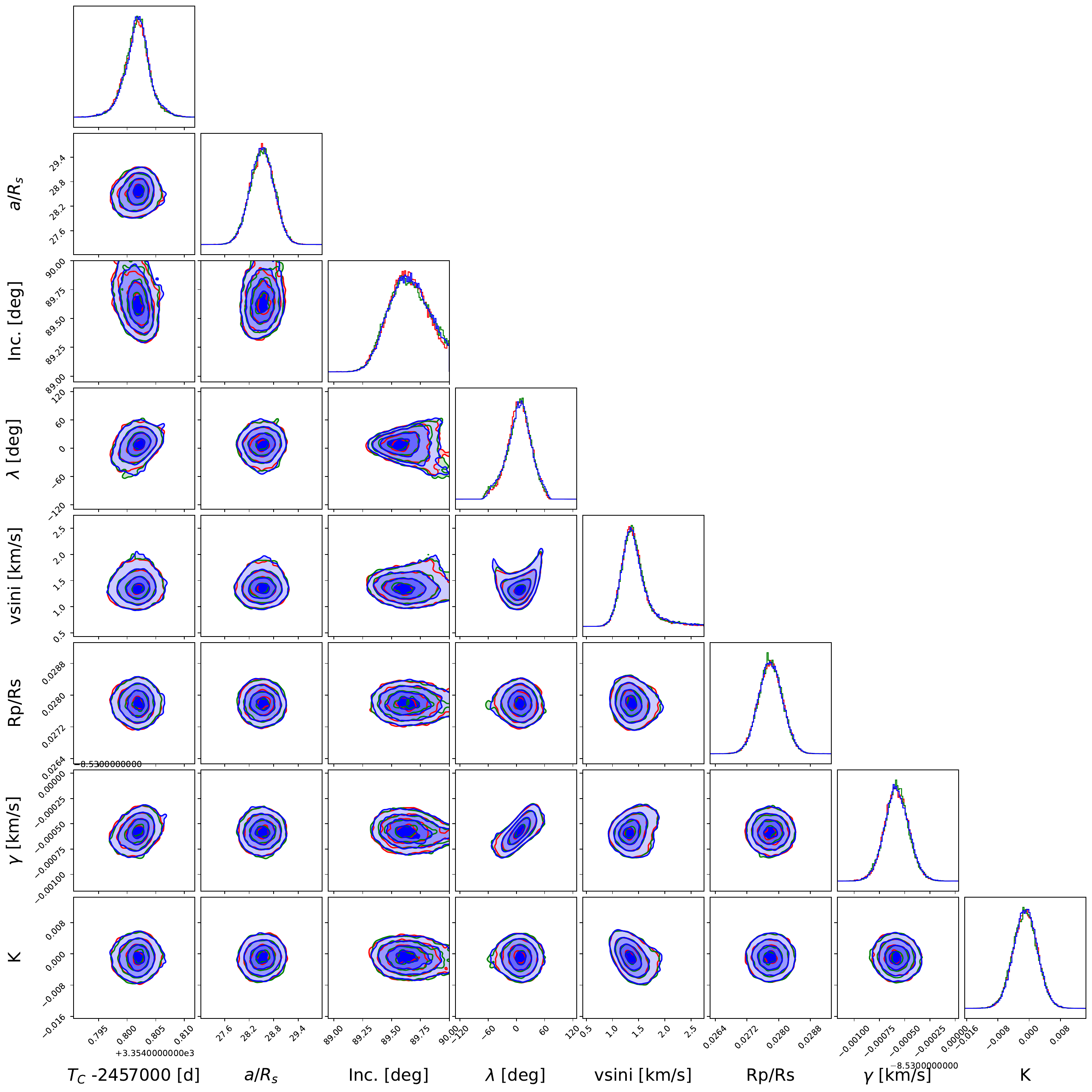}
\caption{Corner plot of the MCMC analysis. Three independent MCMC simulations are shown with different colours.}
\label{f:rmmcmc}
\end{figure*}

\begin{figure*}[ht]
\includegraphics[width=1\textwidth]{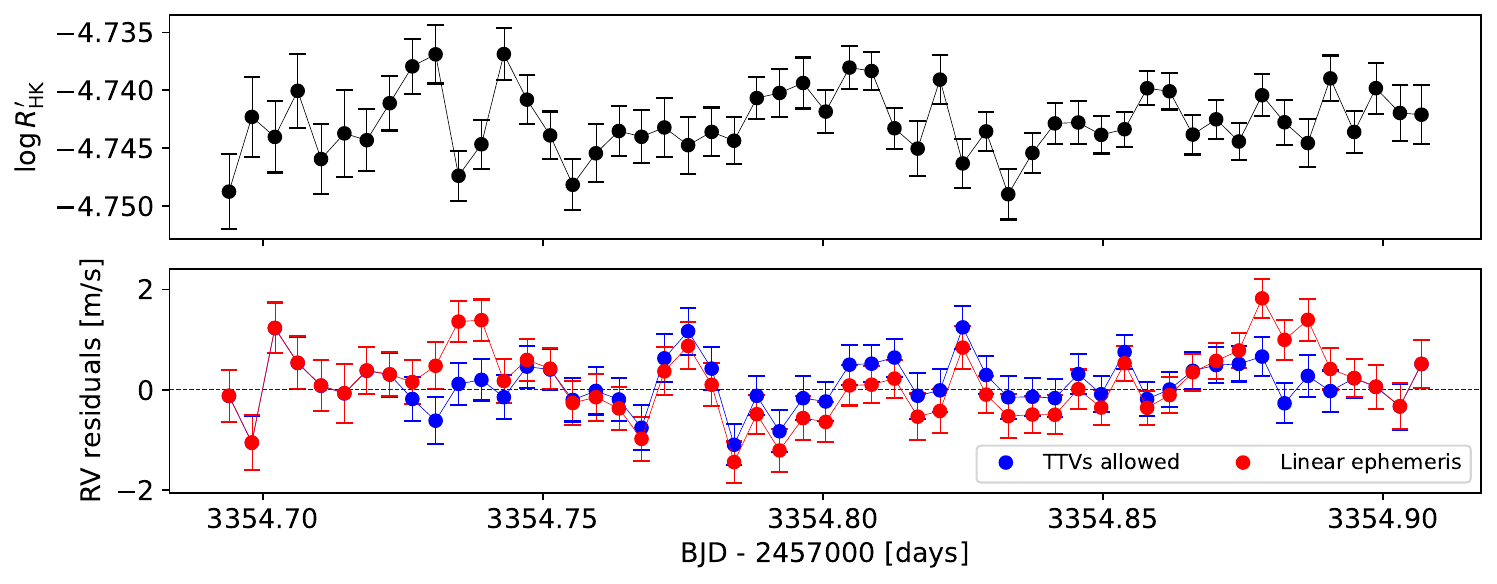}
\caption{(\textit{Top}) Values of the activity index $\log R^\prime_{\mathrm{HK}}$ throughout the night. This index is derived from the Calcium H and K lines and is indicative of chromospheric activity. The low scatter does not support stellar activity significantly influencing obtained RV dataset. (\textit{Bottom}) RV residuals after subtracting our R-M effect model from the observed data (blue). RV residuals after subtracting R-M effect model with linear ephemeris \citep[red,][]{luq23}. The blue model is preferred by the lower standard deviation. }
\label{f:loghk}
\end{figure*}

\end{appendix}

\end{document}